# Modified E91 protocol demonstration with hybrid entanglement photon source

Mikio Fujiwara,[1,*] Ken-ichiro Yoshino,[2] Yoshihiro Nambu,[3] Taro Yamashita,[4] Shigehito Miki,[4] Hirotaka Terai,[4] , Zhen Wang, [4, 5] Morio Toyoshima,[1] Akihisa Tomita,[6] and Masahide Sasaki[1]

[1]*Quantum ICT Laboratory National Institute of Information and Communications Technology (NICT), 4-2-1 Nukui-kita, Koganei, Tokyo 184-8795, Japan*
[2]*Green Platform Research Laboratories, NEC Corporation, 1753 Shimonumabe, Nakahara-ku, Kawasaki 211-8666, Japan*
[3]*Smart Energy Research Laboratories, NEC Corporation, 34 Miyukigaoka, Tsukuba,Ibaraki 305-8501, Japan*
[4]*Nano ICT laboratory, National Institute of Information and Communications Technology (NICT), 588-2 Iwaoka, Kobe 651-2492, Japan*
[5]*Shanghai Center for SuperConductivity, Shanghai Institute of Microsystem and Information Technology, Chinese Academy of Sciences, R3317, 865 Changning Road, Shanghai 200050, PR China*
[6]*Graduate School of Information Science and Technology, Hokkaido University, Kita 14 Nishi 9, Kita-ku, Sapporo, Hokkaido 060-0814, Japan*
[*]*fujiwara@nict.go.jp*

**Abstract:** We report on an experimental demonstration of the modified Ekert 91 protocol of quantum key distribution using a hybrid entanglement source with two different degrees of freedoms, a 1550 nm time-bin qubit and 810 nm polarization qubit. The violation of the Clauser-Horne-Shimony-Holt inequality could be demonstrated for the entanglement between the polarization qubit in free space and the time-bin qubit through 20 km fiber transmission. The secure key rate in our system is estimated 70-150 bps.

**OCIS codes:** (270.5568) Quantum cryptography; (060.5565) Quantum communication.

## 1. Introduction

Data theft by directly tapping installed fibers is becoming a reality these years [1]. Quantum key distribution (QKD) [2] provides a mean to detect such hacking on a physical channel, and allows two users to share random numbers with the unconditional security based on the fundamental laws of physics. Recently QKD systems have been deployed in the field environment [3-5] in parallel with commercialization [6,7]. Moreover, some organizations have started practical use of QKD for their confidential communications [8,9]. Theoretical [10,11] and experimental [12-14] efforts have been made continuously to enhance the performance and the reliability of QKD systems. Decoy-state methods [15,16] for the Bennett and Brassard scheme (BB84) [17] and a derived protocol [18] are widely used for long distance QKD through field installed fibers.

Meanwhile entanglement QKD protocols are studied as next generation technology, which will eventually be combined with quantum repeaters for extending the range of QKD networks [19]. Entanglement based QKD has higher tolerance to side channel attacks than BB84. In addition, it does not need a random number generator for key bits, allowing simpler implementation of the system. Recently, long distance entanglement distribution has been demonstrated, such as over a 300 km fiber [20] and 144 km in free space [21]. Future QKD network would integrate the fiber transmission and free−space transmission to extend the range of applications. The entanglement based QKD would be a favorable option for this purpose because a seamless quantum link with much less security loopholes could be made based on it.

For fiber-based QKD, the time-bin format [22] is widely adopted because of its robustness against polarization mode disturbance. On the other hand, for free space transmission, the polarization format is mostly used. It also allows straightforward implementation of encoding and decoding with high precision and stability by using compact polarizing optical components.

To combine ease of handling a polarization qubit and high-compatibility with fiber transmission of a time-bin qubit, we proposed a hybrid entanglement photon pair source with these degrees of freedoms in different wavelengths and demonstrated entanglement distribution over a hybrid link between a fiber and free space channels [23]. The time-bin qubit was made in a telecom wavelength while the polarization qubit in a near-infrared wavelength. The latter allowed us to use compact silicon avalanche photodiodes (APDs) to detect the polarization qubit efficiently.

In this paper, we present the implementation of a modified Ekert 91(E91) protocol with the hybrid entanglement photon pair source. In E91 protocol [24], three bases are used for testing Bell inequality violation to detect eavesdropping. Recently Acín and co-workers proposed a modified E91 protocol, in which a combination of three bases on one side and two on the other side was used [25], which is simpler than the original E91 protocol where three bases are used on both sides. A. Ling and co-workers demonstrated 1.5 km free−space link and succeeded in generating secure keys with the modified E91 link using a polarization entanglement source [26]. In this paper, we demonstrate the violation of a CHSH-type Bell inequality and the implementation of the modified E91protocol through a 20 km fiber.

## 2. Experimental setup of a modified E91 with a hybrid entanglement photon source

Figures 1(a)-1(c) show the conceptual view of our experiment setup. The hybrid entanglement photon source is constructed of a non-degenerate photon pair source and a format transformer [23] from time encoding to polarization encoding. A 30 mm long periodically poled lithium niobate (PPLN) bulk crystal is quasiphase-matched to generate copolarized entangled photons at 810 and 1550 nm. A 532 nm continuous wave (CW) laser of 130 mW is used as a pump laser. After non-degenerate down conversion, 810 and 1550 nm photons are separated via a dichroic mirror. A 1550 nm photon is sent to Bob via a 22 km dispersion shifted fiber (DSF), and is input to the decoder. The decoder is a two-input and four-output silica-based planer lightwave circuit (PLC) on a silicon substrate [27], as depicted in Fig. 3(c), featured by an asymmetric Mach-Zehnder interferometer (AMZI) with 800 ps time delay. The output photons of the decoder are projected onto vectors |0>, |0>−|1>, |0>+|1>, and |1>, where |0> represents a photon in the first time-bin (having passed through the short arm), and |1> represents a photon in the second time-bin (having passed through the long arm).

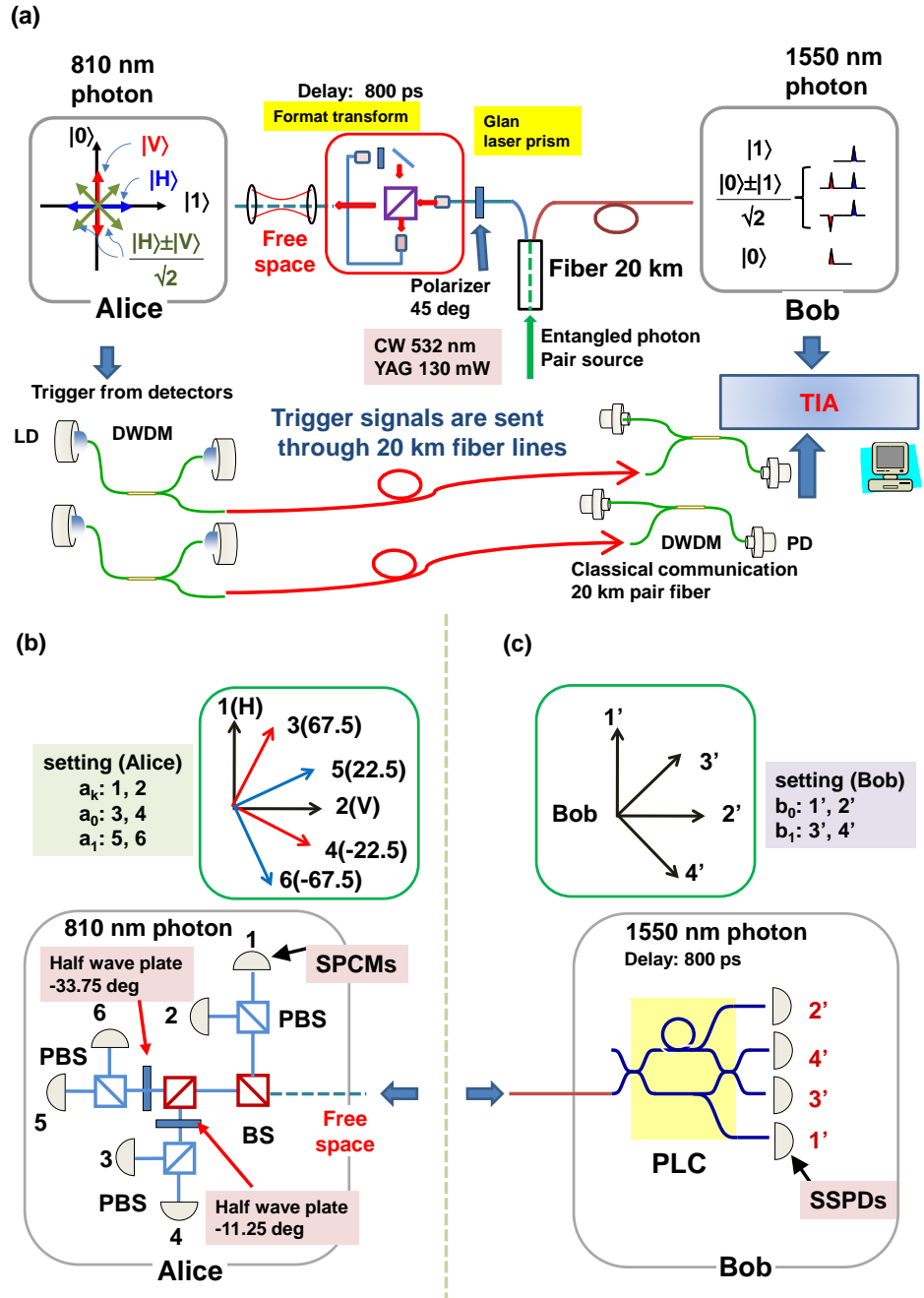

Fig. 1. (a) Conceptual view of the modified E91 protocol QKD system setup.
(b) Receiving set for 810 nm photon (Alice). (c) Decoding set for 1550 nm photon (Bob). PBS: Polarization Beam Splitter.

An 810 nm photon is input to a format-transformer, whose detailed structure and an equivalent optical setup are shown in Figs. 2(a)-2(b). The format-transformer consists of a polarizer set at 45° and a polarization sensitive AMZI constructed of a Glan laser prism, polarization-maintained (PM) fiber, and a mirror [28]. The path difference is set to provide the same delay (800 ps) in Bob's PLC. The polarization qubit thus formed at the 810 nm photon is delivered in free space and coupled to a polarizing decoder circuit at Alice, as shown in Fig. 1(b).

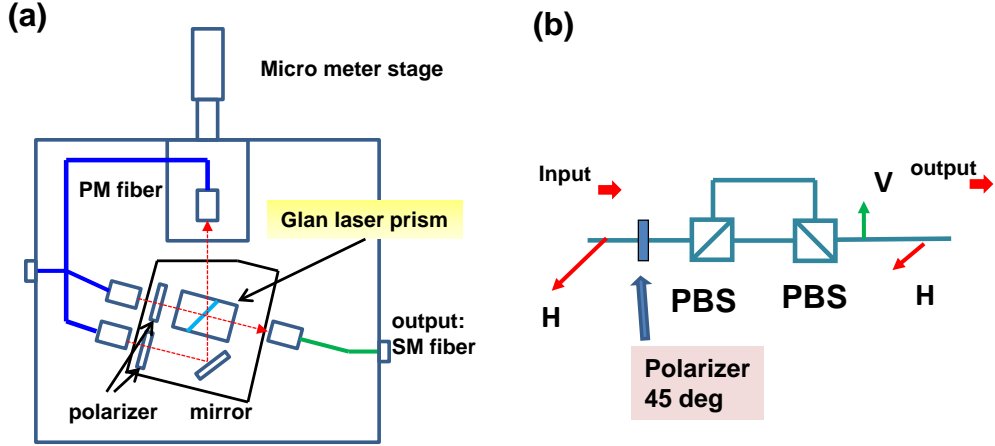

Fig. 2. (a) Conceptual view and (b) equivalent optical setup of the format transformer with 800 ps time delay [28].

The detectors at Alice side are Si based APDs [Perkin Elmer single photon counting modules (SPCMs)]. The detection efficiency and timing jitter of the SPCM are about 55% and 400 ps, respectively. The dark counts of SPCMs are a few kc/s. The detectors at Bob side are superconducting single photon detectors (SSPDs) [29,30]. The SSPDs have the detection efficiencies around 60-70% at the dark count rate of 100 c/s for 1550 nm photons. The coincidence counts are measured by the time interval analyzer (TIA [Hydra harp 400]). In order to compensate time delay of Bob side due to the fiber transmission, we add delay lines for the signals from Alice, as described in the following. Optical pulses synchronized with the SPCM signals are generated using laser diodes. The optical pulses are multiplexed with the trigger signals by dense wavelength division multiplexing (DWDM) filters and transmitted through 20 km single mode (SM) fibers. Those signals are detected by PIN photodiode, and sent to the TIA.

Through the format-transformer, the entanglement state is created in the following form;

$$|\phi\rangle = \frac{1}{\sqrt{2}}\{|H\rangle_A|0\rangle_B + \exp[i\theta(-\tau)-i\theta(0)]|V\rangle_A|1\rangle_B\}, \quad (1)$$

where $|H\rangle$ and $|V\rangle$ represent horizontal and vertical polarization states, respectively. The indices A and B abbreviate Alice and Bob. The relative phase $\theta(t)$ is defined with respect to a reference path length difference $\tau$ between the short and the long arms.

As mentioned in [26], in the modified E91 protocol, the entanglement state written in eq. (1) is measured by a detector unit with the combination of three and two bases shown in Fig. 1(b) and 1(c). The three bases at Alice side are set with polarization angle offsets with 0°, -22.5°, and -67.5°, respectively. The pair of detector setting ($a_k$, $b_0$) is used to generate the raw key. The combinations of other settings are applied to test the violation of the Clauser-Horne-Shimony-Holt (CHSH) [31] inequality $|S|\leq 2$ with

$$S = E(a_0,b_0) + E(a_0,b_1) + E(a_1,b_0) - E(a_1,b_1), \quad (2)$$

where $E$ is the correlation coefficients determined by the coincidence counts between detectors at Alice and Bob. The number of $n_{i,j}$ corresponds to the coincidence count between the detector "$i$" at Alice side and detector "$j$" at Bob side. Correlation coefficients $E$ are given as follows;

$$E(a_0, b_0) = \frac{n_{3,1'} + n_{4,2'} - n_{3,2'} - n_{4,1'}}{n_{3,1'} + n_{4,2'} + n_{3,2'} + n_{4,1'}}$$

$$E(a_0, b_1) = \frac{n_{3,3'} + n_{4,4'} - n_{3,4'} - n_{4,3'}}{n_{3,3'} + n_{4,4'} + n_{3,4'} + n_{4,3'}}$$

$$E(a_1, b_0) = \frac{n_{5,2'} + n_{6,1'} - n_{5,1'} - n_{6,2'}}{n_{5,2'} + n_{6,1'} + n_{5,1'} + n_{6,2'}}$$

$$E(a_1, b_1) = \frac{n_{5,4'} + n_{6,3'} - n_{5,3'} - n_{6,4'}}{n_{5,4'} + n_{6,3'} + n_{5,3'} + n_{6,4'}} .$$

(3)

According to [32], a tight bound on the Holevo information between one of the authorized parties and the eavesdropper is given as;

$$I_{Eve} = h\left(\frac{1 + \sqrt{S^2/4 - 1}}{2}\right)$$

(4)

with the binary entropic function $h(x) = x\log_2 x - (1-x)\log_2(1-x)$. This formula provides an unconditional security bound against the optimal collective attack. To operate this setup as the QKD system, time stamp units are set on both sides, and correlated photons are identified by their time of arrival at each detector. According to the pair of detector ports, coincidence counts are used for generating raw keys or estimating the CHSH-type Bell inequality. After estimating of the eavesdropper knowledge and quantum bit error rate, the secure key is distilled. We estimate the secure key rate by $I_{(A,B)}$-$I_{Eve}$. In the next section, we show the experimental results.

## 3. Experimental results

Figure 3 shows the degree of violation of the CHSH-type Bell inequality as a function of the PLC operation temperature. The phase of the PLC is almost linearly shifted in this region. Coincidence counts are accumulated for 10 s with the time window of 64 ps. The value of $S$ is almost equal when we apply the time window with 128 ps, however, it changes drastically with longer time window. The error bar on the value of $S$ is estimated to be less than 2% using three time measurements. The error bars are smaller than the size of the symbol. The influence of the accidental coincidence counts on both detectors is negligible small due to the narrow time window, so that we don't correct the accidental coincidence counts. The violation of the CHSH inequality is obtained after 20 km fiber transmission with a modified E91 protocol type implementation.

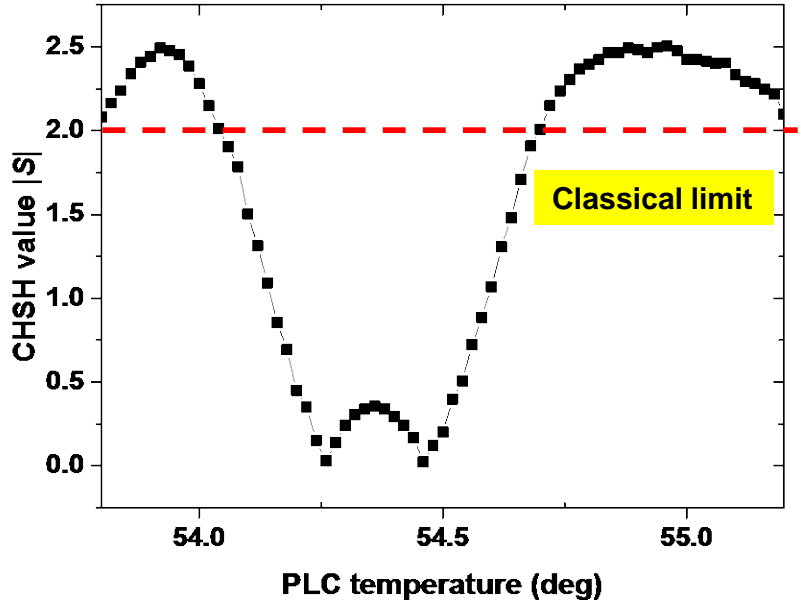

Fig. 3. The degree of violation of a CHSH type Bell inequality as a function of the PLC operation temperature.

To examine the stability of the current implementation, fluctuation of the value of *S* and error rates in raw key are measured over 4 hours and are shown in Fig. 4(a). Although the violation of CHSH inequality holds for 200 minutes, it could not last over 4 hours. The origin of the instability is the drift of the format transformer by temperature fluctuation. Optical path length in the format transformer is influenced by environmental temperature due to the thermal expansion. Unfortunately, the thermal control device is not installed in this transformer yet. In the next generation system, the thermal control in the format transformer is necessary to generate secure keys stably. Nevertheless, the bit error rate is stable during measurement time (180 min) with the average bit error rate is 3.71% shown in Fig. 4(b). Raw keys are generated using the combination of $|H\rangle_A|0\rangle_B$ and $|V\rangle_A|1\rangle_B$, which are insensitive to the phase of photons. The average raw key rate is about 1.5 kbps. The overlap of the time window due to the timing jitter of the SPCM is the other main reason of degradation of the value of *S*. Therefore the improvement of the secure key rate with shortening time delay in the AMZI must be followed by the shortening of the timing jitter in a SPCM.

The average information leakage ($I_{Eve}$) per raw key defined in [32] can be estimated using Eq. (4), and plotted in Fig. 4(c). The error bar in this figure is also smaller than the size of the symbol. The difference of the mutual information between Alice and Bob $I_{(A, B)}$ and the leakage information corresponds to the secure key rate. The mutual information between Alice and Bob $I_{(A, B)}$ estimated by the averaged bit error rate is 0.771 shown with red line in Fig. 4(c). The secure key rate ($I_{(A,B)}-I_{Eve}$) is estimated 70-150 bps with our experimental setup.

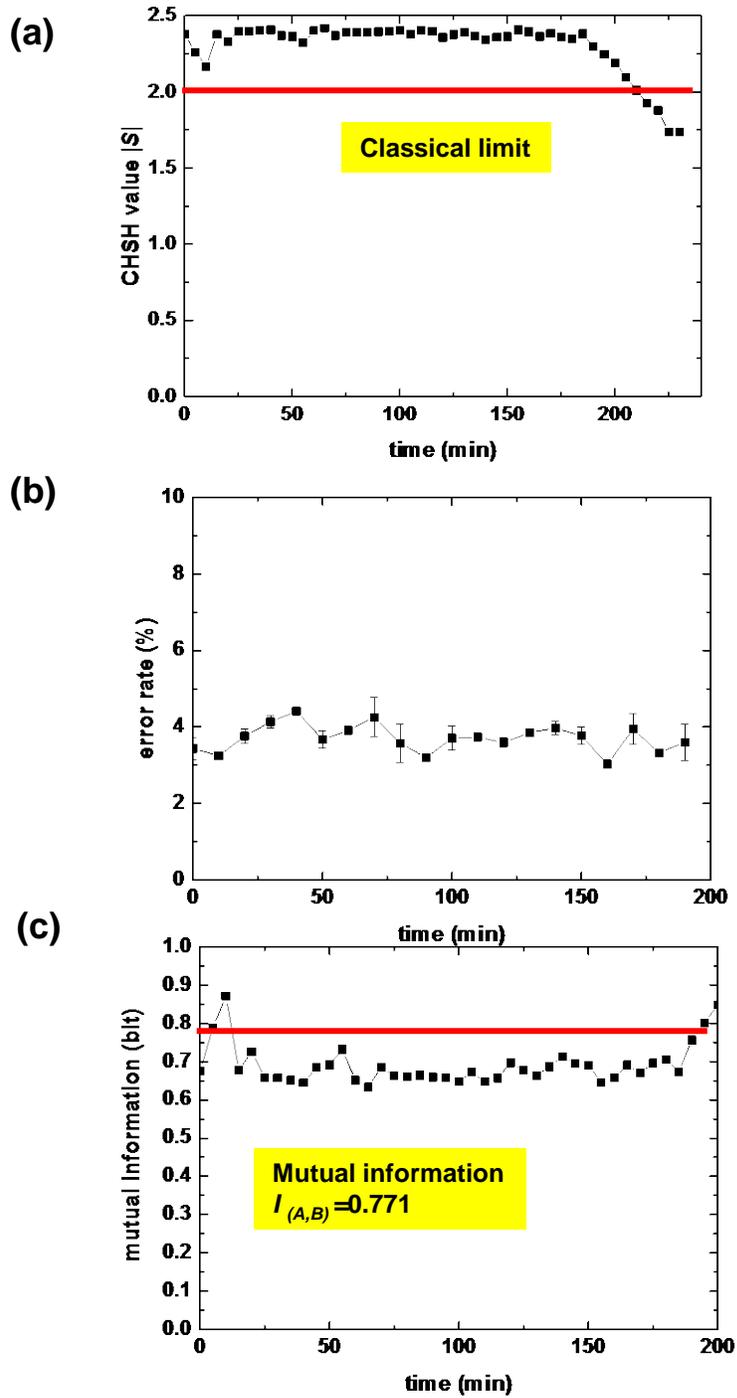

Fig. 4. Experimental result in a stability measurement with implementing a modified E91 protocol. (a) The degree of violation of the CHSH type Bell inequality, (b) the error rate in the raw key, and (c) the unconditional bound obtained with Eq. (4). Red lines in (a) and (c) indicate the classical limit and the mutual information estimated from the average bit error rate, respectively.

## 4. Conclusion

We confirm the entanglement between the polarization qubit in free space and time-bin qubit transmitted through a 20 km fiber with the modified E91 protocol setup. The degree of the violation of the CHSH type Bell inequality and the bit error rate of the raw key indicate that the secure key can be generated with our setup. The key component of this system is the hybrid entanglement photon pair source with the format transformer. It enables us to implement various protocols such as bit commitment or oblivious transfer on the optical fiber. Those protocols provide features different from the QKD function, and proof of principle demonstrations have been carried out with polarization qubit [33,34]. Our hybrid entanglement photon source can integrate the advantages of polarization and time-bin qubits and implement such protocols with one-to-many service system easily. The hybrid entanglement would thus play indispensable roles in secure photonic networks.